\documentclass[aps,prb]{revtex4}
\usepackage{amsmath}
\usepackage{graphicx}

\begin{document}

\title{The Casimir frictional drag force between a SiO$_2$ tip and  a graphene-covered 
SiO$_2$ substrate}

\author{A.I.Volokitin$^{1,2}$\footnote{Corresponding author.
\textit{E-mail address}:alevolokitin@yandex.ru}}
 \affiliation{$^1$Peter Gr\"unberg Institut,
Forschungszentrum J\"ulich, D-52425, Germany}

\affiliation{
$^2$Samara State Technical University, Physical Department, 443100 Samara, Russia}

\begin{abstract}

The possibility of the mechanical  detection of the Casimir friction using  
non-contact force microscope  is discussed.
 On a SiO$_2$ tip 
situated above  a graphene-covered SiO$_2$  substrate will act the frictional drag  force 
mediated  by a fluctuating electromagnetic field produced by a current  in 
the graphene sheet. This friction force will produce 
the bending of the cantilever, which can be measured by  state-of-art 
non-contact force microscope. Both the thermal and quantum contributions to the  
Casimir frictional drag force 
can be studied using this experimental setup. 
This result paves 
the ways for the mechanical detection of the Casimir friction and for the application of the 
frictional drag effect in micro- and nano- electromechanical devices (MEMS and NEMS).
\end{abstract}
\maketitle


\vskip 5mm

\section{Introduction}
All media are surrounded by a fluctuating electromagnetic field due to the thermal and 
quantum fluctuations of the current and charge 
densities inside them. These electromagnetic fluctuations are the cornerstone of the 
Casimir physics which includes the Casimir -- van der Waals forces 
\cite{Casimir1948,Lifshitz1955,Dalvit2011}, the Casimir friction with its limiting 
case —-  quantum friction \cite{VolokitinRMP2007,Dalvit2011},  
and the near-field radiative heat transfer \cite{Van Hove,VolokitinRMP2007}.
The thermal and quantum fluctuation of the current density in one body induces 
the current density in other body; the interaction between these current densities 
is the origin of the Casimir interaction. When two bodies are in relative motion, 
the induced current will lag slightly behind the fluctuating current inducing it, and 
this is the origin of the Casimir friction. At present the Casimir friction is attracting 
a lot of attention due to the fact that it is one of the mechanisms of noncontact
friction between bodies in the absence of direct contact \cite{VolokitinRMP2007}. 
The noncontact friction determines the ultimate limit to which the friction force 
can be reduced and, consequently, also the force fluctuations because they are linked to
friction  via the fluctuation-dissipation theorem. 
The force fluctuations (and hence friction) are important for the ultrasensitive 
force detection.

The Casimir friction  has been studied 
in the configurations plate-plate 
\cite{PendryJPCM1997,VolokitinJPCM1999,VolokitinPRL2003,
VolokitinPRB2003,VolokitinRMP2007,VolokitinPRB2008} and   neutral particle-plate 
\cite{TomassonePRB1997,VolokitinPRB2002,VolokitinRMP2007,DedkovPLA2005,
DedkovJPCM2008,BrevikEntropy2013,KardarPRD2013,BrevikEPJD2014,BartonNJP2010,
HenkelNJP2013,
DalvitPRA2014,VolokitinNJP2014,HenkelJPCM2015}. 
While the predictions of the theory for the Casimir forces were verified in many 
experiments \cite{Dalvit2011}, the detection of the
Casimir friction  is still challenging problem for  experimentalists. However, 
the frictional  drag between quantum wells \cite{GramilaPRL1991,SivanPRL1992} 
and  graphene sheets
\cite{KimPRB2011,GeimNaturePhys2012}, and the current-voltage dependence of 
nonsuspended graphene on the surface of the polar dielectric SiO$_2$ 
\cite{FreitagNanoLett2009}, were accurately described using the
theory of the Casimir friction \cite{VolokitinPRL2011,VolokitinJPCM2001b,
VolokitinEPL2013}. 

\begin{figure}
\includegraphics[width=0.40\textwidth]{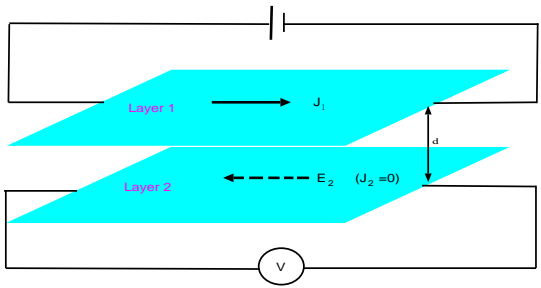}
\caption{\label{fig25} Scheme of experiment for observation of  the drag effect.}
\end{figure}

The frictional drag effect consists in driving an electric current in one
metallic layer and registration of the effect of the frictional
drag of the electrons in a second (parallel) metallic layer (Fig.
\ref{fig25}). Such experiments were predicted by Pogrebinskii \cite
{Pogrebinskii1977} and Price \cite{Price1983} and were performed for 2D
quantum wells \cite{GramilaPRL1991,SivanPRL1992}. In these
experiments, two quantum wells are separated by a dielectric layer
thick enough to prevent electrons from tunneling across it but
allowing interlayer interaction between them. A current of density
$J_2=n_2ev$ is driven through layer \textbf{2} (where $ n_2 $ is
the carrier concentration per unit area in the second layer), see
 Fig.\ref{fig25}. Due to the proximity of the layers, the
interlayer interactions will induce a current in layer \textbf{1}
due to a friction stress $\sigma $ acting on the
electrons in  layer \textbf{1} from layer \textbf{2}. To linear order in the drift velocity $\sigma=\gamma v$ where $\gamma$ is the friction coefficient. If layer
\textbf{1} is an open circuit, an electric field $E_1$ will
develop in the layer whose influence cancels the frictional stress
$\sigma$ between the layers. Thus the frictional stress $ \sigma
=\gamma v$ must equal the induced stress $n_1eE_1$ so that
\begin{equation}
\gamma = n_1eE_1/v=n_2n_1e^2E_1/J_2=n_1n_2e^2\rho_{D},
\end{equation}
 where the drag resistivity $\rho_{D}=E_1/J_2=\gamma/n_1n_2e^2$ is defined
as the ratio of the induced electric field in the first layer to
the driving current density in the second layer. The
transresistivity is often interpreted in terms of a drag rate
which, in analogy with the Drude model, is defined by
$\tau_D^{-1}=\rho_{12}n_2e^2/m^*= \gamma/n_1m^*$.
Frictional drag between graphene sheets was measured recently in Refs. \cite{KimPRB2011,GeimNaturePhys2012}. This study has fueled the recent theoretical investigations of frictional drag between graphene sheets  mediated by a fluctuating Coulomb field \cite{Tse2007,Katsnelson2011,Peres2011,Hwang2011,Narozhny2012,Katsnelson2012,Amorin2012} (see Ref. \cite{NarozhnyRMP2016} for recent review).
Most of this work focused on interlayer Coulomb interaction, the most
obvious coupling mechanism and the one considered in the original
theoretical papers \cite{Pogrebinskii1977}, though the contributions due to an
exchange of phonons between the layers have also been considered \cite
{GramilaPRB1993}.
The most widely
used approach to study the  drag effect is based on the Boltzmann
equation  and the Kubo formalism
\cite{NarozhnyRMP2016}. In the Fermi-liquid regime, $k_BT\ll \epsilon_F$,  and in the limit of strong screening, $k_{TF}d\gg 1$, the drag resistivity for both the 2D-quantum wells and  the graphene sheets is given by \cite{Zheng,Tso,Jauho,Kamenev,Tse2007}
\begin{equation}
\rho_D=\frac{\gamma}{(ne)^2}=\frac{h}{e^2}\frac{\pi\zeta(3)}{32}\left(\frac{k_BT}{\epsilon_F}\right)^2\frac{1}{(k_Fd)^2}\frac{1}{(k_{TF}d)^2},  \label{2Dfricdrag}
\end{equation}
where for the 2D-quantum wells $k_{TF}=2a_0^{-1}/\varepsilon $ is the single-layer
Tomas-Fermi screening wavevector, $a_0=\hbar^2 /m^{*}e^2$, and for the graphene sheets $k_{TF}=4e^2k_F/\varepsilon\hbar v_F$, $m^*$ is the electron effective mass, $k_F$ is the Fermi wave number, $\epsilon_F$ is the Fermi energy, $v_F$ is the Fermi velocity, $d$ is the separation between 2D layers, $\varepsilon$ is the dielectric constant for the surrounding medium. 

\begin{figure}
\includegraphics[width=0.40\textwidth]{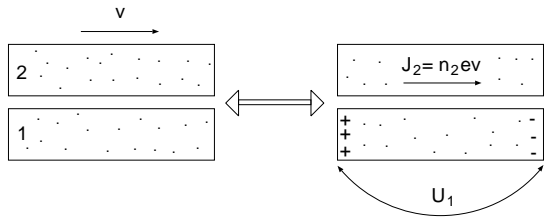}
\caption{ Two ways of studying of the Casimir friction.
Left: a metallic block sliding relative to the metallic
substrate with the velocity $v$. An electronic frictional stress
will act on the block (and on the substrate). Right: the shear
stress $\sigma$ can be measured if instead of sliding the upper
block, a voltage $U_2$ is applied to the block resulting in a
drift motion of the conduction electrons (velocity $v$). The
resulting frictional stress $\sigma$ on the substrate
electrons will generate a voltage difference $U_1$ (proportional to $\sigma$
) as indicated in the figure. Both approaches are equivalent if the conduction electrons are in the Fermi-liqid state and it is possible to neglect  scattering of the free carries by lattice. 
 \label{fig24}}
\end{figure}

The close connection of the Casimir friction with frictional drag  effect  is illustrated in Fig. \ref{fig24}.
Application of the theory of the Casimir friction to these experiment is based on the assumption that the Fermi-liquid theory is valid for electrons in graphene. In this case the graphene with the drift motion of electrons with the drift velocity $v_{Drift}$ produces the same fluctuating electromagnetic field and has the same reflection amplitudes as graphene moving with velocity $v=v_{Drift}$. The hydrodynamic model for electrons in a medium was used in Ref.\cite{ShapiroPRB2010} to calculate  the fluctuating electromagnetic field produced by the medium in the presence of the drift motion of the electrons.  
  A theory of the Casimir friction was used for the description of the frictional drag effect in the 2D quantum wells in Ref.\cite{VolokitinJPCM2001b} and in the graphene double layer in \cite{VolokitinEPL2013}. Eq. (\ref{2Dfricdrag}) was reproduced using such approach. However, the theory of the Casimir friction, in contrast to the theory of the Coulomb drag, includes the contribution  not only from the fluctuating Coulomb field due to the charge fluctuations but also the contribution from the electromagnetic field produced by the transverse current density fluctuations.

In one experiment  (see \cite{GramilaPRL1991}) the drift velocity $v_{Drift}\sim 10^2$ m/s. According to the theory of the Casimir friction \cite{VolokitinPRL2011,VolokitinJPCM2001b,
VolokitinEPL2013}, at such velocities the thermal fluctuation give the dominant contribution to the friction, and the theoretical predictions are in an agreement with experiment. However, at large drift velocities the Casimir friction is dominated by the contribution from quantum fluctuations which are not included in the theory of the Coulomb drag. Using the theory of the Casimir friction in Ref. \cite{VolokitinPRL2011} it was shown that the current-electric field dependence of nonsuspended graphene on the SiO$_2$ substrate can be explained by quantum friction between electrons in graphene and surface phonon polaritons in  SiO$_2$.

In the frictional drag experiment the electric field induced by the the Casimir friction force 
is measured. For the graphene sheet situated nearby the polar dielectric 
substrate the Casimir friction force between the charge free carries in graphene and the 
surface phonon polaritons in dielectric gives rise to the change of the resistivity of 
graphene which also can be measured. So far the Casimir friction was detected only using 
the electrical effects, which it produces. Thus, the frictional drag effect can only 
be observed between the two 2D conducting structures and the electrical transport 
in graphene can only be measured for nonsuspended graphene when the heat conductance 
between graphene and underlying dielectric is high.

\begin{figure}[tbp]
\includegraphics[width=0.80\textwidth]{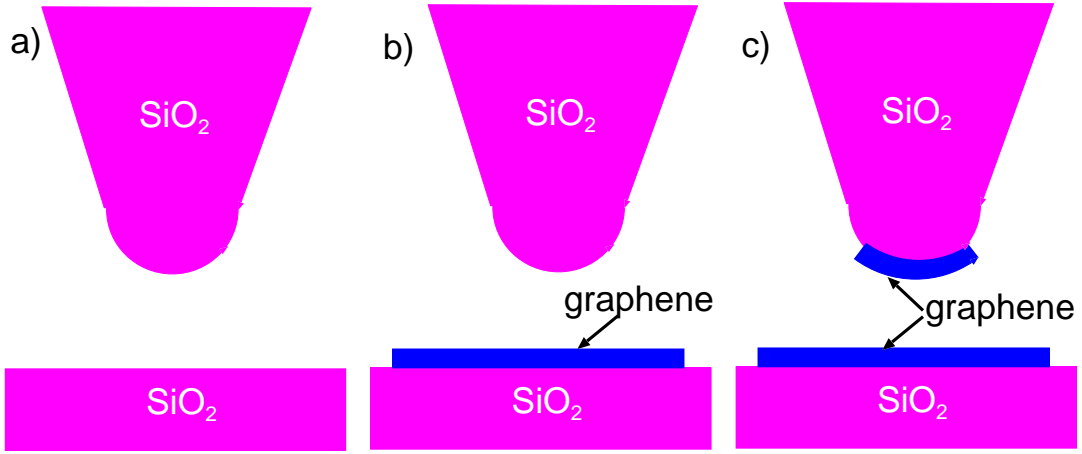}
\caption{The different configurations for the observation of the Casimir friction
 using non-contact force microscope: a) A SiO$_2$ tip and a SiO$_2$ substrate (DD); b) A SiO$_2$ tip and a graphene-
covered SiO$_2$ substrate (DGD); c)  A graphene-
covered SiO$_2$ tip and a graphene-
covered SiO$_2$ substrate (DGGD).}
\label{MechDet}
\end{figure}

In this paper the possibility of the mechanical detection of 
 the Casimir friction using non-contact atomic force 
microscope (AFM) is considered. This topic we also discussed recently in Ref.\cite{VolokitinJETPLett2016}. The schemes for the experimental setups are shown on 
Fig.\ref{MechDet}. On Fig.\ref{MechDet}a a SiO$_2$ tip and a SiO$_2$  
substrate have  clean surfaces (DD structure). On  Fig.\ref{MechDet}b  a SiO$_2$  
substrate is covered by graphene and a SiO$_2$ tip has clean surface (DGD structure), and on 
Fig.\ref{MechDet}c both surfaces of the tip and the dielectric are covered by graphene (DGGD structure).

This paper is organised as follows. In Sec. \ref{PP} the Casimir friction is considered in the plate-plate configuration for the SiO$_2$ plates and the graphene-covered 
SiO$_2$ plates. To linear order in the relative velocity $v$ the friction stress between plates $f =\gamma v$ where the friction coefficient $\gamma$ was calculated analytically in the resonant and off-resonant approximations.  In Sec. \ref{TP} the Casimir friction in the tip-plate configurations is calculated using the proximity force approximation. In Sec. \ref{Drag} the Casimir frictional drag force acting on the tip due to the drift motion of the electrons in the graphene sheet on the SiO$_2$ plate is calculated. The summary of the obtained results is given in Sec. \ref{Summary}. 

\section{Casimir friction in the plate-plate configuration \label{PP}}

According to the theory of the Casimir friction, the 
frictional stress between two plates in the parallel relative motion with 
the velocity $v$ along the $\hat{x}$-axis and with the $\hat{z}$-axis normal to the plate is 
given by the $xz$-component of the Maxwell stress tensor: 
 $f=\sigma _{xz}=f_{T}+f_{0}$, where at $d\ll\lambda_T=c\hbar/k_BT$ and 
$v\ll c$ the contributions from thermal ($f_{T}$) and quantum ($f_{0}$) fluctuations 
are given by \cite{VolokitinJPCM1999,VolokitinRMP2007,VolokitinPRB2008,VolokitinPRL2011}

\[
f_{T} =\frac \hbar {\pi ^3}\int_{0 }^\infty dq_y\int_0^\infty
dq_xq_xe^{-2qd}\Bigg \{ \int_0^\infty d\omega \Bigg(
\frac{\mathrm{Im}R_{1}(\omega)\mathrm{Im}R_{2}(\omega^+) }{\mid
1-e^{-2 q d}R_{1}(\omega)R_{2}(\omega^+)\mid ^2}\times
 [n_1(\omega )-n_2(\omega^+)]+(1\leftrightarrow 2)\Bigg )
\]
\begin{equation}
 -\int_0^{q_xv}d\omega \Bigg(\frac{\mathrm{Im}R_{1}(\omega)\mathrm{Im}
R_{2}(\omega^-)} {\mid 1-e^{-2qd}R_{1}(\omega)R_{2}(\omega^-)\mid
^2} n_1(\omega) +(1\leftrightarrow 2)\Bigg )\Bigg \}, \label{FrictionT}
\end{equation}
\begin{equation}
f_{0} =-\frac \hbar {2\pi ^3}\int_{0 }^\infty dq_y\int_0^\infty
dq_xq_xe^{-2qd}\int_0^{q_xv}d\omega \Bigg(\frac{\mathrm{Im}R_{1}(\omega)\mathrm{Im}
R_{2}(\omega^-)} {\mid 1-e^{-2qd}R_{1}(\omega)R_{2}(\omega^-)\mid
^2}  +(1\leftrightarrow 2)\Bigg ).\label{Friction0}
\end{equation}
where   $\omega^{\pm}=\omega \pm q_xv$, $R_{ip}$ is the reflection amplitude for the $p$-
polarized electromagnetic wave for the plate $i$, 
$n_i(\omega)=[\exp(\omega/k_BT_i)-1]^{-1}$. The symbol $(1 \leftrightarrow 2)$ denotes 
the terms that are obtained from
the preceding terms by permutation of $1$ and $2$.
The reflection amplitude for the dielectric for $d<c/(\omega_0|\epsilon_d|)$ 
\begin{equation}
R_{d}=\frac{\epsilon _{d}-1}{\epsilon _{d}+1},
 \label{refcoef}
\end{equation}
where $\epsilon _{d}$ and $\omega_0$ are  the dielectric function and the characteristic frequency for 
dielectric.
The dielectric function of amorphous SiO$_2$ can be described
using an oscillator model\cite{Chen2007APL}
\begin{equation}
\epsilon(\omega) =
\epsilon_{\infty}+\sum_{j=1}^2\frac{\sigma_j}{\omega_{0,j}^2-\omega^2-i\omega\gamma_j},
\end{equation}
where parameters $\omega_{0,j}$, $\gamma_j$ and $\sigma_j$ were
obtained by fitting the actual $\epsilon$ for SiO$_2$ to the above
equation, and are given by $\epsilon_{\infty}=2.0014$,
$\sigma_1=4.4767\times10^{27}$s$^{-2}$, $\omega_{0,1}=8.6732\times
10^{13}$s$^{-1}$, $\gamma_1=3.3026\times 10^{12}$s$^{-1}$,
$\sigma_2=2.3584\times10^{28}$s$^{-2}$, $\omega_{0,2}=2.0219\times
10^{14}$s$^{-1}$, and $\gamma_2=8.3983\times 10^{12}$s$^{-1}$. 
For a graphene-covered SiO$_2$ substrate 
the reflection amplitude $R_{dg}$  can be 
expressed through the reflection amplitudes for the clean substrate 
surface $R_d$ given
by Eq. (\ref{refcoef}) 
and 
for  isolated graphene given by \cite{VolokitinJPCM2001b} 
\begin{equation}
R_g=(\varepsilon_g-1)/\varepsilon_g,
\end{equation} 
where the dielectric function of graphene 
\begin{equation}
\varepsilon_g=1 + \frac{2\pi i q\sigma_l}{\omega},
\end{equation}
where $\sigma_l$ is the longitudinal conductivity of the sheet. To get the expression for $R_{dg}$ we assume that the graphene layer is located at 
$z=0$ and the dielectric at $z<-a$. The electric field can be written in the form
\begin{equation}
\mathbf{E}(\mathbf{q},\omega ,z)=  \left\{
\begin{array}{rl}
 R_{dp}\hat{n}_p^+e^{-qz} +
\hat{n}_p^-e^{qz}
, &\text{for}\, \,x>0 \\
v_{p}\hat{n}_p^{+}e^{-qz} + w_{p}\hat{n}_p^{-}e^{qz}, &\text{for}\, \,-a<z<0 
\end{array}\right.,
\label{1surf6}
\end{equation}
where  $\hat{n}_p^{\pm}=(\mp i\mathbf{q},q)$, $a$ is the separation between graphene and the dielectric. From continuity of the tangential component of the 
electric field at $z=0$ follows: $1-R_{dg}=w_p-v_p$. The second boundary condition at $z=0$ follows from the requirement that at $-a<z<0$ the amplitude of the outgoing wave 
is equal to the amplitude of the reflected wave plus the amplitude of the transmitted wave: $w_p=R_gv_p + 1-R_g$, and  at $z=-a$ the amplitude of the outgoing wave 
is equal to the amplitude of the reflected wave: $v_p=e^{-2qd}R_dw_p$. 
From these boundary conditions for  $qa \ll 1$ follows
 \begin{equation}
R_{dg}=1-\frac{(1-R_d)(1-R_g)}{1-R_dR_g} = \frac{\epsilon_d - 1 +2(\varepsilon_g - 1)}
{\epsilon_d + 1 +2(\varepsilon_g - 1)}.
\label{refdg}
\end{equation}
In the study below we used the dielectric function of graphene, which was
calculated  within the random-phase approximation (RPA)
\cite{Wunsch2006,Hwang2007}.  The dielectric function
is an analytical function in the upper half-space
of the complex $\omega$-plane:
\begin{equation}
\varepsilon_0(\omega,q)=1+\frac{4k_Fe^2}{\hbar
v_Fq}-\frac{e^2q}{2\hbar \sqrt{\omega^2-v_F^2q^2}}\Bigg \{G\Bigg
(\frac{\omega+2v_Fk_F}{v_Fq}\Bigg )- G\Bigg
(\frac{\omega-2v_Fk_F}{v_Fq}\Bigg )-i\pi \Bigg \},
\end{equation}
where
\begin{equation}
G(x)=x\sqrt{x^2-1} - \ln(x+\sqrt{x^2-1}),
\end{equation}
where the Fermi wave vector $k_F=(\pi n)^{1/2}$, $n$ is the
concentration of charge carriers, the Fermi energy
$\epsilon_F=\hbar v_Fk_F$,  $v_F\approx 10^6$ m/s is the Fermi velocity. 

To linear order in the velocity $v$ the friction force $f=\gamma v$ where 
at $T_1=T_2=T$, the 
friction coefficient  
\begin{equation}
\gamma =\frac {\hbar^2} {8\pi ^2k_BT}\int_0^\infty
\frac{d\omega}{ \sinh^2\left(\frac{\hbar \omega}{2k_BT}\right)}
\int_0^\infty dq\,q^3e^{-2qd}\frac {\mathrm{Im}R_{1p}\mathrm{Im}R_{2p}}{\left|
1-e^{-2qd}R_{1p}R_{2p}\right| ^2}.
\label{parallel6}
\end{equation}
The reflection amplitude for a dielectric given by Eq.(\ref{refcoef}) has the resonance at the condition $\epsilon^{\prime}_d(\omega_s)=-1$ where $\epsilon_d^\prime$ is the real part 
of the dielectric function $\epsilon_d$. This condition determines the frequency $\omega_s$ of the surface phonon polariton mode. Close to the resonance the reflection amplitude can 
be written in the form \cite{VolokitinPRB2004,VolokitinRMP2007}
\begin{equation}
R_d\approx -\frac{\omega_a}{\omega - \omega_s +i\eta}.\label{res}
\end{equation}
Close to the resonance the transmission coefficient in Eq. (\ref{parallel6}) for two identical dielectrics can be written in the form
\begin{equation}
t=\frac {(\mathrm{(Im}R_de^{-qd})^2}{\left|
1-e^{-2qd}R_{d}R_{d}\right| ^2}\approx \frac {(\omega_a\eta e^{-qd})^2}{[(\omega-\omega_+)^2+\eta^2][(\omega-\omega_-)^2+\eta^2]},
\label{trcoef}
\end{equation}
where 
\[
\omega_{\pm}=\omega_s\pm \omega_ae^{-qd}. 
\]
Using Eq. (\ref{trcoef}) in Eq. (\ref{parallel6}) gives the resonant contribution to the friction coefficient
\[
\gamma_{res} \approx \frac {\hbar^2\eta} {4\pi k_BT
 \sinh^2\left(\frac{\hbar \omega_s}{2k_BT}\right)}
\int_0^\infty dq\,q^3\frac {(Be^{-qd})^2}{(Be^{-qd})^2+1} \approx \frac {\hbar^2\eta} {4\pi k_BT
 \sinh^2\left(\frac{\hbar \omega_a}{2k_BT}\right)}
\int_0^{q_c} dq\,q^3
\]
\begin{equation}
=\frac {\hbar^2\eta q_c^4} {16\pi k_BT
 \sinh^2\left(\frac{\hbar \omega_s}{2k_BT}\right)},
\label{res}
\end{equation}
where it was assumed that $B=\omega_a/\eta  > 1$, $\omega_s\gg \omega_a\mathrm{exp}(-qd)$, $q_c=\mathrm{ln}B/d$. At small frequencies far from the resonance ($\omega\ll\omega_s$) $t\approx (\omega \mathrm{exp}(-qd)/\omega^*)^2$ and the off-resonant contribution to the friction coefficient 
\begin{equation}
\gamma_{offres}\approx=\frac {\hbar} {16d^4}
\left(\frac{k_BT}{\hbar \omega^*}\right)^2.
\label{offres}
\end{equation}
For SiO$_2$ $\omega_s=9,6\cdot 10^{13}$s$^{-1}$, $\omega_a=4.5\cdot 10^{12}$s$^{-1}$, $\eta=1.7\cdot 10^{12}$s$^{-1}$, $\omega^*=2.3\cdot 10^{16}$s$^{-1}$. With these 
parameters Eqs. (\ref{res}) and (\ref{offres}) at $T=300$K and $d=1$nm give $\omega_{res}=3.5\cdot 10^{-2}$kgs$^{-1}$m$^{-2}$ and  $\omega_{0ffres}=1.8\cdot 10^{-5}$kgs$^{-1}$m$^{-2}$. 

Another resonance is possible in the condition of the anomalous Doppler effect when $q_xv=2\omega_s$ \cite{Jacob.J.Opt.2014,Jacob.Opt.Exp.2014}. At this resonant condition, taking into account that 
$R(-\omega)=R^*(\omega)$,    at 
$q_y=0$ the denominators in the 
integrands in Eqs. (\ref{FrictionT}) and (\ref{Friction0}) contain the factor
\begin{equation}
1-|R(\omega_s)|^2e^{-\frac{4\omega_s d}{v}}.
\end{equation}
At the resonance $|R(\omega_s)|$ can be larger than unity thus the denominator is equal to zero at 
\begin{equation}
v_c=\frac{2\omega_s d}{\mathrm{ln}|R(\omega_s)|}
\end{equation}
what means that above the threshold velocity at $v>v_c$ the friction force for two identical plates is divergent. The origin of this divergence is 
related with the creation above the threshold velocity $v_c$ of the non-dissipative collective eigenmode for two identical plates even in the case when the surface phonon polariton modes for 
the isolated surfaces are dissipative. Above the threshold velocity $v_c$ the amplitude of this mode increases infinitely with time what gives rise to the divergence of the dissipation and friction \cite{SilveirinhaNJP2014}. 

Close to the resonance at $\omega^{\prime}=q_xv-\omega= \omega_s$ the critical velocity $v_c=2\omega_sd/\mathrm{ln}(\omega_a/\eta$ and the reflection amplitude can be written in the form
\begin{equation}
R_d(\omega-q_xv)\approx -\frac{\omega_a}{ \omega_s-\omega^{\prime} -i\eta}.\label{antires}.
\end{equation}
Close to the resonance at $\omega\approx \omega^{\prime}\approx\omega_s$  the transmission coefficient in Eq. (\ref{parallel6}) for two identical dielectrics can be written in the form
\begin{equation}
t=\frac {\mathrm{Im}R_d(\omega)\mathrm{Im}R_d(\omega-q_xv)e^{-2qd}}{\left|
1-e^{-2qd}R_{d}(\omega)R_{d}(\omega-q_xv)\right| ^2}\approx \frac {(\omega_a e^{-qd})^2}{(\omega-\omega^{\prime})^2+\left[\frac{(\omega-\omega_s)(\omega^{\prime}-\omega_s)+\eta^2-
\omega_s^2\mathrm{exp}(-2qd)}{\eta}\right]^2},
\label{trcoefanti}
\end{equation}

Using Eq. (\ref{trcoefanti}) in Eq. (\ref{Friction0}) gives the resonant 
contribution to the friction coefficient from $\omega\approx \omega_s\approx q_xv-\omega_s$
\begin{equation}
f_{friction}\approx -\frac {\hbar\eta^2
\mathrm{ln}^{2.5}(\omega_a/\eta)}
{2\pi \omega_s d^3 }\mathrm{ln}\frac{v_c-v}{v_c}.
\label{frresanti}
\end{equation}
The divergent factor $\mathrm{ln}(v_c-v)/v_c$ in Eq. (\ref{frresanti}) can be also 
written in the form 
$\mathrm{ln}(d-d_0)/d_0$ where $d_0= v\mathrm{ln}(\omega_a/\eta)/(2\omega_s)$. 
Thus at the given velocity $v$ the friction force diverges at $d<d_0$.

\section{Casimir friction in the tip-plate configuration \label{TP}}

An atomic force microscope tip with the radius of curvature $R\gg d$, 
at a distance $d$ above a flat
sample surface, can be
approximated by a sphere with radius $R$. In this case the friction force
 between the tip and the plane surface can be estimated using the 
approximate method of Derjaguin \cite{Derjaguin1934}, later
called the proximity force  approximation (PFA) \cite{PFA1977}. According to 
this method, the friction force in
the gap between two smooth curved surfaces at short
separation can be calculated approximately as a sum of
forces between pairs of small parallel plates corresponding
to the curved geometry of the gap.  Specifically, the sphere-plane  friction force
 is given by 
\begin{equation}
F=2\pi \int_0^R d\rho \rho f(z(\rho )),  \label{approx}
\end{equation}
where   $z(\rho )=d+R-\sqrt{R^2-\rho^2}$ denotes the tip-surface distance as a function of
the distance $\rho $ from the tip symmetry axis, and the friction force
per unit area $f(z(\rho ))$ is determined in the plate-plate configuration. This
scheme was proposed in \cite{Derjaguin1934,Hartmann} for the calculation of the
conservative van der Waals interaction; in this case the error is
not larger than 5-10\% in an atomic force application, and 25\% in
the worst case situation \cite{Apell}. We assume that the same
scheme is also valid for the calculation of the Casimir friction. However, as it was discussed in Sec. \ref{PP} for two identical plates 
the 
Casimir friction diverges at the velocities above the Cherenkov threshold velocity. However, For the different plate the friction force can be finite 
even above the threshold velocity. For the SiO$_2$ the threshold velocity $v_c\approx 2\cdot 10^5$m/s. Thus in the present study the numerical calculations are performed 
at the velocities below the threshold velocity when one can assume that the PFA gives  sufficiently accurate estimation of the Casimir friction 
including the quantum friction.  
During the last few years, the most general method available for calculating
both Casimir force and radiative heat transfer between many bodies of arbitrary shapes, materials,
temperatures and separations was obtained which expresses the
Casimir  force and radiative heat transfer  in terms of the scattering matrices of individual bodies \cite{Dalvit2011}.
Specifically, the numerically exact solution for the
near-field radiative heat  transfer between a sphere and an infinite plane was first performed using the scattering matrix
approach. In principle the same approach can be used for the calculation of the Casimir friction. 
We
assume that the tip has a paraboloid shape given [in cylindrical
coordinates ($z,\rho $)] by the formula: $z=d+\rho ^2/2R$, where
$d$ is the distance between the tip and the flat surface. If
\begin{equation}
f=\frac C{\left( d+\rho ^2/2R\right) ^n},
\end{equation}
we get
\begin{equation}
F=\frac{2\pi R}{n-1}\frac C{d^{n-1}}=\frac{2\pi Rd}{n-1}f(d)\equiv
A_{ \mathrm{eff}}S(d),  \label{approx1}
\end{equation}
where $A_{\mathrm{eff}}=2\pi Rd/(n-1)$ is the effective surface area.
In a more general
case one must use numerical integration to obtain the friction  force.
\begin{figure}[tbp]
\includegraphics[width=0.40\textwidth]{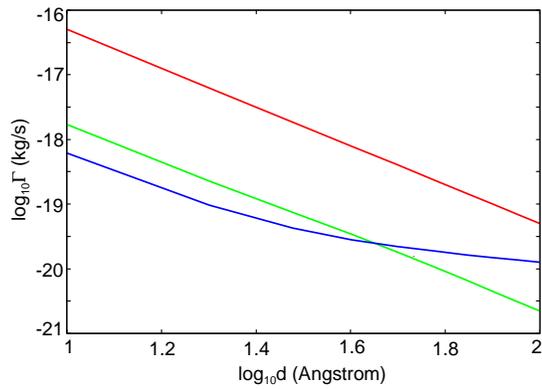}
\caption{The dependence of the friction coefficient $\Gamma$ for the spherical tip 
with the radius of the curvature $R=1\mu$m on the 
separation $d$ between the tip and the substrate for the  different configurations 
at $T=300$K. The red, green and blue curves represent the results for the configurations 
(a) SiO$_2$-SiO$_2$ (Fig.\ref{MechDet}a),  (b) SiO$_2$+graphene-SiO$_2$ (Fig.\ref{MechDet}b), and (c) 
SiO$_2$+graphene-SiO$_2$+graphene (Fig.\ref{MechDet}c), respectively. The charge carriers concentration for graphene 
$n=10^{16}$m$^{-2}$}.  
\label{FrCoef}
\end{figure}

The friction coefficient in the tip-plane configuration $\Gamma$ can be obtained from 
Eq.(\ref{parallel6}) using the proximity force approximation. In an 
experiment $\Gamma$ is determined   by measuring  the quality factor
 of the cantilever 
vibration parallel to the substrate surface\cite{MeyerElements2015}. At present can only 
be detected the friction coefficient in the range $10^{-12}-10^{-13}$kg/s. Fig.\ref{FrCoef} shows the dependence of 
the friction coefficient on the separation between a tip and a substrate surface 
for the different configurations. The red, green and blue curves represent the results 
for the configurations 
(a) SiO$_2$-SiO$_2$ (DD) (Fig.\ref{MechDet}a),  (b) SiO$_2$+graphene-SiO$_2$ (DGD) 
(Fig.\ref{MechDet}b), and (c) SiO$_2$+graphene - SiO$_2$+graphene (DGGD)
(Fig.\ref{MechDet}c), respectively. The friction coefficient in these configurations 
is below $10^{-16}$kg/s thus it can not be tested by the modern experimental setup. 
However,  it has been predicted in Ref.\cite{VolokitinPRB2006a}, that for the some 
configurations involving adsorbates the Casimir friction coefficient can be large enough 
to be measured by state-of-art non-contact force microscope. 

\begin{figure}[tbp]
\includegraphics[width=0.40\textwidth]{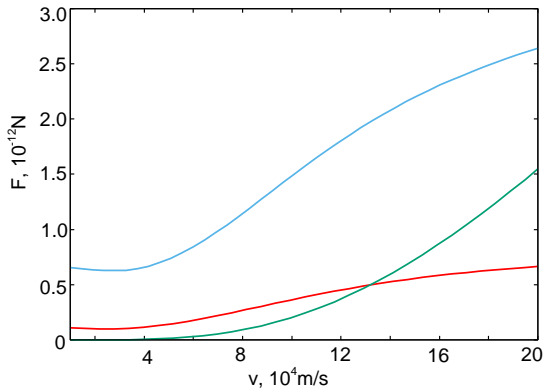}
\caption{The dependence of the different contributions to the friction force  
 on the 
relative sliding velocity  $v$ between a tip and a substrate  for the SiO$_2$-SiO$_2$ 
configuration (see Fig. \ref{MechDet}a). The blue and red lines show the thermal contributions at $T=600$K and $T=300$K, 
respectively. The green line shows the quantum contribution ($T=0$K). The radius of the curvature of the tip $R=1\mu$m. The separation 
between the tip and substrate $d=1$nm.}
 \label{DD}
\end{figure}

\begin{figure}[tbp]
\includegraphics[width=0.40\textwidth]{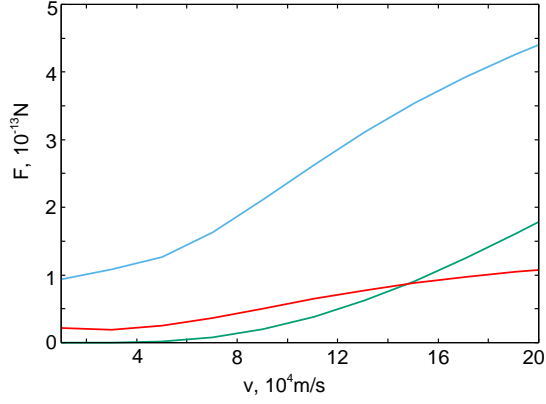}
\caption{The same as on Fig. \ref{DD} but for  the SiO$_2$+graphene-SiO$_2$ 
configuration (see Fig. \ref{MechDet}b).}
 \label{DGD}
\end{figure}

During the cantilever vibration the 
velocity 
of the AFM tip does not exceed $1$m/s. However, the Casimir friction force can be strongly 
enhanced at the large relative sliding velocity. This friction force  can be detected 
if it  produces sufficiently large bending of the cantilever. Fig. \ref{DD} shows 
the dependence of the friction force, acting on the tip with the radius 
of the curvature $R=1\mu$m, on the relative velocity $v$ between the tip 
and substrate for the SiO$_2$-SiO$_2$ 
configuration (see Fig.\ref{MechDet}a) at the separation 
$d=1$nm and  for the different  temperatures. The friction force $F=F_0+
F_T$ where $F_0$ is the contribution from quantum fluctuations which exist even at 
$T=0$K (this friction is denoted as quantum friction \cite{PendryJPCM1997},  
$F_{friction}(T=0$K)$=F_0)$ and $F_T$ 
is the contribution from the thermal fluctuations which exist only at finite temperature. 
The thermal contribution dominates for $v<k_BTd/\hbar$ and quantum contribution dominates 
for $v>k_BTd/\hbar$. On Fig.\ref{DD}  $F>10^{-12}$N at $v>10^5$m/c. In the modern 
experiment\cite{NanoLett2015}
the spring constant of the cantilever are between $k_0=30$ and 
$k_0=50\mu$N/m. The friction force $\approx 10^{-12}$N will produce the displacement 
of the tip of the order $10^2$nm which can be easily detected. However, at present there
is  no  experimental setup with the relative sliding velocity between the tip and 
substrate $\sim 10^5$m/s. Fig. \ref{DGD} shows the friction force  for the SiO$_2$+graphene-SiO$_2$ 
configuration (see Fig.\ref{MechDet}b) which is one order of the magnitude smaller the friction 
force for the SiO$_2$-SiO$_2$ 
configuration (see Fig. \ref{DD}).

\section{Casaimir friction drag between a tip and a graphene covered SiO$_2$ plate \label{Drag}}

An alternative method for the detection of the Casimir friction is possible for the 
SiO$_2$+graphene - SiO$_2$ configuration (see Fig.\ref{MechDet}b). For this configuration 
inducing current in a graphene sheet with the drift velocity of the free charge 
carriers $v_{Drift}$ will produce 
the fluctuating electromagnetic field which is similar to the electromagnetic field 
due to the mechanical motion of the sheet with the velocity $v=v_{Drift}$ 
\cite{VolokitinJPCM2001b,
VolokitinRMP2007,VolokitinPRL2011,VolokitinEPL2013}.  Due to the high mobility of the charge 
carriers 
in graphene, in a high electric field electrons (or holes)  can move with very high 
velocities (up to $10^6$ m/s). The drift motion  of charge carries in graphene will 
result in a modification of dielectric properties  of graphene  
due to  the Doppler effect \cite{PendryJPCM1997}. The reflection amplitude for the graphene 
sheet with induced  current density   is determined by the reflection 
amplitude $R_g^{\prime}$ in the reference frame co-moving with the drift motion of the 
charge carriers in graphene: $R_g^{\prime}=R_g(\omega^{-})$, where $R_g(\omega)$ 
is the reflection amplitude amplitude in the rest reference frame  
of the graphene sheet without current, $\omega^{-}=\omega-q_xv$, $v=v_{Drift}$. 

In the vacuum
gap between two plates in the configuration  SiO$_2$+graphene - SiO$_2$  the electric 
field $\mathbf{E}(\mathbf{q},\omega,z)$
can be written in the form \cite{VolokitinPRB2008}
\begin{equation}
\mathbf{E}(\mathbf{q},\omega ,z)= 
v_p\hat{n}_p^+e^{-qz} +
w_p\hat{n}_p^-e^{qz},
  \label{gapone}
\end{equation}
where $\hat{n}_p^{\pm}=(\mp i\mathbf{q},q)$, 
\begin{equation}
v_p=\frac{E_{dg}^{f\prime} + R_{dg}^{\prime}E_d^fe^{-qd}}{1-e^{-2qd}R_{dg}^{\prime}R_d},\,
w_p=\frac{R_dE_{dg}^{f\prime}e^{-2qd} + E_d^fe^{-qd}}{1-e^{-2qd}R_{dg}^{\prime}R_d},
\end{equation}
where $E_{dg}^{f\prime}$ and $E_d$ are the amplitudes of the fluctuating 
electric fields created on the surfaces of plates  by the charge density fluctuations inside the SiO$_2$+graphene  and 
SiO$_2$ plates, respectively, and where, in the presence of the  drift motion of the 
free charge carriers  
in a graphene sheet with the drift velocity $v_{Drift}$, 
the reflection amplitude 
$R_{dg}^{\prime}$ is given by Eq.(\ref{refdg}) with $R_{g}$  replaced on 
$R_{g}^{\prime}=R_{g}(\omega^{-})$
\begin{equation}
R_{dg}^{\prime}=1-\frac{(1-R_d)(1-R_g^{\prime})}{1-R_dR_g^{\prime}} = 
\frac{\epsilon_d - 1 +2(\varepsilon_g^{\prime} - 1)}
{\epsilon_d + 1 +2(\varepsilon_g^{\prime} - 1)},
\label{refdgpr}
\end{equation}
where $\varepsilon_g^{\prime}=\varepsilon_g(\omega^{-})$. To get the expression for $E^f_{dg}$, resulting from the interference of the electric fields created 
by the SiO$_2$ plate and the graphene sheet, we assume that the graphene layer is located at 
$z=0$ and the dielectric at $z<-a$.
 The electric field can be written in the form
\begin{equation}
\mathbf{E}(\mathbf{q},\omega ,z)=  \left\{
\begin{array}{rl}
 E^f_{dp}\hat{n}_p^+e^{-qz}
, &\text{for}\, \,x>0 \\
v_{p}\hat{n}_p^{+}e^{-qz} + w_{p}\hat{n}_p^{-}e^{qz}, &\text{for}\, \,-a<z<0 
\end{array}\right.,
\label{1surf6}
\end{equation}
where  $\hat{n}_p^{\pm}=(\mp i\mathbf{q},q)$, $a$ is the separation between graphene and the dielectric. From continuity of the tangential component of the 
electric field at $z=0$ follows: $E^f_{dg}=v_p - w_p$. The second boundary condition at $z=0$ follows from the requirement that at $-a<z<0$ the amplitude of the outgoing wave 
is equal to the amplitude of the reflected wave plus the amplitude of the  wave emitted by 
the graphene sheet due to the charge fluctuations inside the sheet: $w_p=R_gv_p -E^f_g$, 
and  at $z=-a$ the amplitude of the outgoing wave 
is equal to the amplitude of the reflected wave plus the amplitude of the  wave 
emitted  by the SiO$_2$ plate: $v_p=e^{-2qd}R_dw_p +E_d^f$. 
From these boundary conditions for  $qa \ll 1$ follows
\begin{equation}
E_{dg}^{f\prime}=\frac{E_{d}^{f}(1-R_{g}^{\prime})+E_{g}^{f\prime}(1-R_d)}
{1-R_dR_{g}^{\prime}}=\frac{E_{d}^{f}(\epsilon_d + 1)+2E_{g}^{f\prime}\varepsilon_g^{\prime}}
{\epsilon_d + 1 +2(\varepsilon_g^{\prime} - 1)},
\label{elfieldgd}
\end{equation}
where $E_{d}^{f}$ and $E_{g}^{f\prime}$ are the electric fields   created by the charge density fluctuations in   the SiO$_2$ 
plate and in the graphene sheet with the drift motion of the charge carriers, respectively.
According to the
general theory of the fluctuating electromagnetic field \cite{VolokitinRMP2007}
the spectral density of the fluctuations for the electric field:

\begin{equation}
<|E^{f}_i|^2>_{\omega}=\frac{2\hbar }
{q} \left(n_i(\omega)+\frac{1}{2}\right)\mathrm{Im}R_i,  \label{ImR}
\end{equation}
where $<...>$ denote statistical average over the random field, $i=d,g$. 
The frictional stress   $f_x $ acting on the surface of the SiO$_2$ plate 
is determined by   $xz$-component
of Maxwell's stress tensor $\sigma _{ij}$, calculated at
$z=+0$:
\begin{equation}
f_x=\sigma_{xz} =\frac 1{8\pi }\int_{-\infty }^{+\infty }d\omega
[\langle E_{z}E_{x}^{*}\rangle+\langle E_{z}^{*}E_{x}\rangle
\rangle ]_{z=+0}=
 2\mathrm{Im}\int_0^{\infty }\frac{d\omega}{2\pi} \int
\frac{d^2q}{(2\pi)^2}\frac{q_x}q 
\left\langle w_p^*v_p \right\rangle, 
\label{stressdgd}
\end{equation}
where the symbol $\langle ..\rangle $ denotes statistical averaging
on the random fields  $\mathbf{E}_d^f$ and $\mathbf{E}_g^f$. Using 
Eqs. (\ref{gapone})-(\ref{elfieldgd}) in Eq.(\ref{stressdgd}), after averaging 
over the random 
electric fields in Eq.(\ref{stressdgd}) using Eq.(\ref{ImR}), we get
\[
f_x=\frac{\hbar}{2\pi^3}\int_0^{\infty}d\omega\int d^2q\,q_x\frac{\mathrm{Im}R_d}
{\left|1-e^{-2qd}\mathrm{Im}R_d \mathrm{Im}R_{dg}^{\prime}\right|^2}
\Big(\mathrm{Im}R_{dg}^{\prime}[n_g(\omega^{-})-n_d(\omega)]
\]
\begin{equation}
+\frac{2\mathrm{Im}\epsilon_d}{|\epsilon_d+1+2(\varepsilon_g^{-}-1)|^2}
[n_g(\omega)-n_g(\omega^{-})]\Big).
\label{stresdgd}\   
\end{equation}
In Eq. (\ref{stresdgd})  the factors $n_g$ and 
$n_d$    are calculated at the temperatures
 $T_g$ and $T_d$ for the SiO$_2$+graphene and SiO$_2$ 
plates, respectively. The contribution to the friction from the quantum fluctuations 
can be obtained from Eq. (\ref{stresdgd}) at $T_g=T_d=0$ K
\begin{equation}
f_x(T=0\mbox{K})=-\frac{2\hbar}{\pi^3}\int d^2q\,q_x\int_0^{q_xv}d\omega
\frac{\mathrm{Im}R_d\mathrm{Im}\varepsilon_g^{\prime}}
{\left|1-e^{-2qd}\mathrm{Im}R_d \mathrm{Im}R_{dg}^{\prime}\right|^2
|\epsilon_d+1+2(\varepsilon_g^{\prime}-1)|^2}
\label{stresdgd0}\   
\end{equation}

\begin{figure}[tbp]
\includegraphics[width=0.40\textwidth]{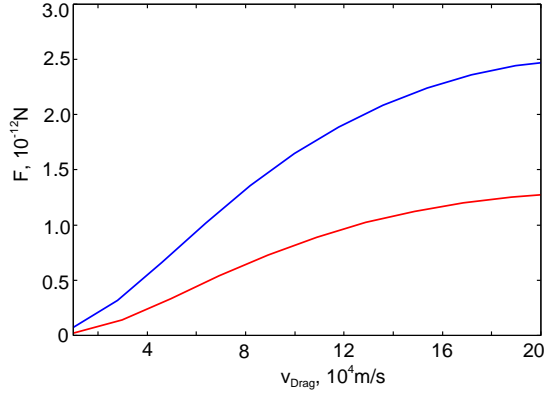}
\caption{The same as on Fig. \ref{DGD} but for the dependence of the thermal contributions to the friction force 
 on the drift velocity $v_{Drift}$
of the charge carriers in graphene. The quantum friction (not shown) is negligible in the range of the velocities shown on the figure.}
 \label{DGDDrag}
\end{figure} 

Fig.\ref{DGDDrag} shows the dependence of the frictional drag force acting on the SiO$_2$ tip
in the SiO$_2$+graphene-SiO$_2$ configuration on the drift velocity $v_{Drift}$ of electrons 
in the graphene sheet. For $v_{Drift}>10^5$m/s the friction force is above $10^{-12}$N and 
can be measured by  state-of-art non-contact force microscope.  It is important to note that 
in the contrast to the SiO$_2$-SiO$_2$ configuration, where the bending of the cantilever 
due to the Casimir friction can only be detected for very large relative sliding velocity 
between a tip and a substrate, in the SiO$_2$+graphene-SiO$_2$ configuration the friction 
of the same order can be obtained inducing the current density in the graphene sheet by 
high electric field what can be easily obtained using the modern experimental setup.

\section{Summary \label{Summary}} 
A current in the graphene sheet produces a fluctuating 
electromagnetic field which is similar to the field produced by the moving sheet. In 
a high electric field  electrons in non suspended graphene on the SiO$_2$ substrate can 
move with sufficiently large drift velocity (above $10^6$m/s \,\cite{FreitagNanoLett2009}) 
to produce the frictional drag force acting on a tip which can be measured by  
state-of-art non-contact force microscope. Both the thermal and 
quantum contributions to the 
Casimir friction can be detected using this experimental setup.  These results pave 
the ways for the mechanical detection of the Casimir friction and for the application of the 
frictional drag effect in micro- and nano- electromechanical devices (MEMS and NEMS).

The study was supported by  
the Russian Foundation for Basic Research (Grant No. 16-02-00059-a).


\vskip 0.5cm


\begin{thebibliography}{999}

\bibitem{Casimir1948} H.B.G. Casimir,  Proc. K. Ned. Akad. Wet. \textbf{51}, 793 (1948).

\bibitem{Lifshitz1955} E.M. Lifshitz, Zh. Eksp. Teor. Fiz. \textbf{29}, 94 
(1955) [Engl. Trnsl. 1956 Sov. Phys.-JETP \textbf{2}, 73 (1956)].


\bibitem{Dalvit2011} \textit{Casimir Physics.} Ed. by D.A.R. Dalvit, P. Milonni, 
D. Roberts and F. da Rose (Springer, Berlin 2011).




\bibitem{VolokitinRMP2007}  A.I. Volokitin  and  B.N.J. Persson, Rev.Mod.Phys.
\textbf{79}, 1291 (2007).

\bibitem{Van Hove}  D.Polder and M.Van Hove, Phys. Rev. B \textbf{4}, 3303,
(1971).


\bibitem{PendryJPCM1997}  J.B. Pendry,    J. Phys.: Condens.Matter   
\textbf{9},  10301 (1997).



\bibitem{VolokitinJPCM1999} A.I. Volokitin  and  B.N.J. Persson, 
J. Phys.: Condens.Matter   \textbf{11},  345 (1999).

\bibitem{VolokitinPRL2003}  A. I. Volokitin and B. N. J. Persson, Phys. Rev. Lett.
\textbf{91}, 106101 (2003).

\bibitem{VolokitinPRB2003}  A. I. Volokitin and B. N. J. Persson, Phys. Rev. B,
\textbf{68}, 155420 (2003).


\bibitem{VolokitinPRB2008}  A.I. Volokitin  and  B.N.J. Persson,  Phys. Rev. B  
\textbf{ 78}, 155437 (2008).


\bibitem{TomassonePRB1997}  M. S. Tomassone and A. Widom, Phys. Rev. B \textbf{56}, 4938 (1997)

\bibitem{VolokitinPRB2002}  A. I. Volokitin and B. N. J. Persson, Phys. Rev. B \textbf{65}, 115419 (2002)

\bibitem{DedkovPLA2005} G. V. Dedkov and A. A. Kyasov, Phys. Lett. A \textbf{339}, 212 (2005).


\bibitem{DedkovJPCM2008} G. V. Dedkov  and A. A. Kyasov,   \textit{J.Phys.:Condens.Matter} \textbf{20}, 354006 (2008).

\bibitem{BartonNJP2010} G. Barton,  New J. Phys. \textbf{12}, 113045 (2010).



\bibitem{BrevikEntropy2013} J. S. H{\o}ye and I. Brevik, Entropy {\bf 15}, 3045 (2013).

\bibitem{BrevikEPJD2014} J. S. H{\o}ye and I. Brevik, Eur. Phys. J. D {\bf 68}, 61 (2014).

\bibitem{KardarPRD2013} M. F. Maghrebi, , R. Golestanian and M. Kardar, Phys. Rev. D {\bf 87}, 025016 (2013).

\bibitem{DalvitPRA2014} F. Intravaia, R. O. Behunin and D. A. R. Dalvit, Phys. Rev. A {\bf 89}, 050101(R) (2014).

\bibitem{VolokitinNJP2014} A.I. Volokitin  and  B.N.J. Persson, New J. Phys. \textbf{16}, 118001 (2014).

\bibitem{HenkelNJP2013} G. Pieplow  and C. Henkel,   New J. Phys. \textbf{15}, 023027 (2013).

\bibitem{HenkelJPCM2015} G. Pieplow  and C. Henkel,   J. Phys.: Condens. Matter \textbf{27}, 214001 (2015).



\bibitem{GramilaPRL1991}   T.J. Gramila,   J.P. Eisenstein,   A.H. MacDonald,
L.N. Pfeiffer, and K. W. West,  Phys. Rev. Lett. \textbf{66}, 1216
(1991).

\bibitem{SivanPRL1992}   U. Sivan,   P.M. Solomon, and   H. Shtrikman,  Phys. Rev.
Lett.  \textbf{68, }1196 (1992).

\bibitem{GramilaPRB1993}  T.J. Gramila, J.P. Eisenstein, A.H. MacDonald, L.N.
Pfeiffer, and K.W. West, Phys. Rev. B \textbf{47}, 12957 (1993)


\bibitem{KimPRB2011} S.Kim, I.Jo, J.Nah, Z.Yao, S.K.Banerjee  and E.Tutuc, Phys. Rev. B \textbf{83} 161401 (2011).

\bibitem{GeimNaturePhys2012} R.V.Gorbachev, A.K.Geim, M.I.Katsnelson, K.S.Novoselov, T.Tudorovskiy, T.V.Grigorieva, A.H.MacDonald, K.Watanabe,
T.Taniguchi  and L.P.Ponamarenko  Nature Phys. \textbf{8} 896 (2012).

bibitem{Katsnelson2011} M.I.Katsnelson, \textit{Phys. Rev.B} \textbf{84} 041407(R) (2011).

\bibitem{Peres2011} N.M.R.Peres, J.M.R.Lopes des Santos and A.H.Castro Neto, Europhys. Lett \textbf{95} 18001 (2011).

\bibitem{Hwang2011} E.H.Hwang, R.Sensarma and S.DasSarma, Phys. Rev. B \textbf{84} 245441 (2011).

\bibitem{Narozhny2012} B.N.Narozhny, M.Titov, I.V.Gornyi  and P.M.Ostrovsky, Phys. Rev. B \textbf{85} 195421 (2012).

\bibitem{Katsnelson2012} M.Carrega, T.Tudorovskiy, A.Principi, M.I.Katsnelson  and M.Polini M 2012  New J. Phys. \textbf{14} 063033 (2012).

\bibitem{Amorin2012} B.Amorin  and N.M.R.Peres,  J. Phys.:Condens. Matter. \textbf{24} 335602 (2012).

\bibitem{FreitagNanoLett2009} M. Freitag,  M. Steiner,  Y. Martin,
 V. Perebeinos, Z. Chen,  J.C. Tsang, and  P. Avouris,  Nano
Lett.  \textbf{9}, 1883 (2009).



\bibitem{VolokitinJPCM2001b} A.I.Volokitin and  B.N.J. Persson, J.Phys.:
Condens. Matter  \textbf{13}, 859 (2001).

\bibitem{VolokitinPRL2011} A.I. Volokitin  and  B.N.J. Persson, Phys. Rev. Lett. 
\textbf{106}, 094502 (2011).



\bibitem{VolokitinEPL2013} A.I.Volokitin and  B.N.J. Persson, EPL
  \textbf{103}, 24002(2013).

\bibitem{Pogrebinskii1977}   M.B. Pogrebinskii,  Fiz.Tekh.Poluprov.  \textbf{11}, 637 (1977)
 [Sov.Phys. Semicond. \textbf{11}, 372 (1977)].

\bibitem{Price1983}   P. J. Price,  Physica B+C \textbf{117},750 (1983).

\bibitem{Tse2007} W.K.Tse, BenYu-Kuang.Hu and S.DasSarma,  Phys. Rev.B \textbf{76} 081401 (2007).


\bibitem{Katsnelson2011} M.I.Katsnelson, \textit{Phys. Rev.B} \textbf{84} 041407(R) (2011).

\bibitem{Peres2011} N.M.R.Peres, J.M.R.Lopes des Santos and A.H.Castro Neto, Europhys. Lett \textbf{95} 18001 (2011).

\bibitem{Hwang2011} E.H.Hwang, R.Sensarma and S.DasSarma, Phys. Rev. B \textbf{84} 245441 (2011).

\bibitem{Narozhny2012} B.N.Narozhny, M.Titov, I.V.Gornyi  and P.M.Ostrovsky, Phys. Rev. B \textbf{85} 195421 (2012).

\bibitem{Katsnelson2012} M.Carrega, T.Tudorovskiy, A.Principi, M.I.Katsnelson  and M.Polini M 2012  New J. Phys. \textbf{14} 063033 (2012).

\bibitem{Amorin2012} B.Amorin  and N.M.R.Peres,  J. Phys.:Condens. Matter. \textbf{24} 025003 (2016).


\bibitem{NarozhnyRMP2016} B.N.Narozhny  and A. Levchenko, Rev. Mod. Pays.  \textbf{88} 195421 (2012).

\bibitem{Zheng}  L.Zheng and A.H.MacDonald, Phys. Rev. B \textbf{48}, 8203
(1993).

\bibitem{Tso}  H.C.Tso and P.Vasilopoulos, Phys.Rev. B \textbf{45}, 1333
(1992).

\bibitem{Jauho}  A.-P.Jauho and H.Smith, Phys.Rev. B \textbf{47}, 4420 (1993).



\bibitem{Kamenev}  A.Kamenev and Y.Oreg, Phys.Rev. B \textbf{52}, 7516 (1995).

\bibitem{ShapiroPRB2010}  B.Shapiro, Phys. Rev. B \textbf{82}, 075205 (2010).

\bibitem{VolokitinJETPLett2016} A.I.Volokitin, JETP Lett.,  \textbf{104}, 504 (2016).



\bibitem{Chen2007APL}
D.Z.A. Chen, R. Hamam, M. Soljacic, J.D. Joannopoulos and G. Chen,
Appl. Phys. Lett. {\bf 90}, 181921 (2007).

\bibitem{Wunsch2006} B. Wunscvh,  T. Stauber,   F. Sols, and F. Guinea, New J.Phys.  
\textbf{8},318 (2006).


\bibitem{Hwang2007}  E.H. Hwang, S.Das Sarma, Phys. Rev. B \textbf{75}, 205418 (2007).

\bibitem{VolokitinPRB2004}  A.I. Volokitin  and  B.N.J. Persson,  Phys. Rev. B  
\textbf{ 69}, 045417 (2004).

\bibitem{Jacob.J.Opt.2014} Y.Guo and Z.Jacob, J.Opt. \textbf{16}, 114023 (2014).


\bibitem{Jacob.Opt.Exp.2014} Y.Guo and Z.Jacob, Opt. Express. \textbf{22}, 21 (2014).

\bibitem{SilveirinhaNJP2014}  M. G. Silveirinha,  New J. Phys., \textbf{16},  063011 (2014).



\bibitem{Derjaguin1934} B. Derjaguin,   Kolloid-Z. \textbf{69}, 155 (1934).

\bibitem{PFA1977} Blocki, J., J. Randrup, W. J. Swiatecki, and C. F. Tsang, 
Ann.Phys. (N.Y) \textbf{105}, 427 (1977).

\bibitem{Hartmann}  U. Hartmann, Phys. Rev. B \textbf{42}, 1541 (1990);
\textbf{43}, 2404 (1991).

\bibitem{Apell}  P.Johansson and P.Apell, Phys. Rev. B \textbf{56}, 4159
(1997).


\bibitem{MeyerElements2015} E. Gnecco and  E. Meyer,  
\textit{Elements of friction theory and nanotribology}, ed.
by E. Gnecco and E. Meyer (Cambridge University Press 2015).

\bibitem{VolokitinPRB2006a}  A. I. Volokitin and B. N. J. Persson, and H. Ueba, 
Phys. Rev. B \textbf{73}, 165423 (2006).

\bibitem{NanoLett2015} A. Mehlin, F. Xue, D. Liang, H. F. Du, M. J. Stolt, 
S. Jin, M. L. Tian, and M. Poggio, Nano Lett.  \textbf{15}, 4839 (2015).




\end{thebibliography}
\end{document}